\documentclass{PoS}
\usepackage{citesort}
\def\eq{\begin{eqnarray}}
\def\en{\end{eqnarray}}

\title{Partially twisted boundary conditions for scalar mesons}

\ShortTitle{Partially twisted boundary conditions for scalar mesons}

\author{\speaker{Akaki Rusetsky}%
\\
        Helmholtz-Institut fuer Strahlen- und Kernphysik (Theorie) and\\ Bethe Center for Theoretical Physics, Universitaet Bonn, D-53115 Bonn, Germany\\
        E-mail: \email{rusetsky@hiskp.uni-bonn.de}}

\author{Dimitri Agadjanov\\
        Helmholtz-Institut fuer Strahlen- und Kernphysik (Theorie),\\ Bethe Center for Theoretical Physics, Universitaet Bonn, D-53115 Bonn, Germany and\\
St. Andrew the First-Called Georgian University of the Patriarchate of Georgia,\\ Chavchavadze Ave., 53a, 0162, Tbilisi, Georgia\\
        E-mail: \email{dagadjanov@hiskp.uni-bonn.de}}

\author{Ulf-G. Mei{\ss}ner\\
        Helmholtz-Institut fuer Strahlen- und Kernphysik (Theorie),\\ 
Bethe Center for Theoretical Physics, Universitaet Bonn, D-53115 Bonn, Germany and\\
Institute for Advanced Simulation (IAS-4), Institut f\"ur Kernphysik 
(IKP-3) and\\ J\"ulich Center for Hadron Physics,
Forschungszentrum J\"ulich D-52425, J\"ulich, Germany\\
        E-mail: \email{meissner@hiskp.uni-bonn.de}}

\abstract{The possibility of imposing partially twisted boundary conditions 
in the lattice study of the resonance states is investigated by using the
effective field theory (EFT) methods. In particular, it is demonstrated that -- in
certain cases -- it is possible to use partial twisting even in the presence
of the quark annihilation diagrams. This talk is mainly based on our recent
work~\cite{ournew}, which provides substantially more details and discussion.}

\FullConference{31st International Symposium on Lattice Field Theory LATTICE 2013\\
		 July 29 - August 3, 2013\\
		 Mainz, Germany}

\begin{document}

\section{Introduction}
Recently, 
it has been argued that the use of twisted boundary
conditions~\cite{twisted,Sachrajda,Chen} may prove
useful for the extraction of the parameters of the resonances from the lattice
QCD data~\cite{Lage-scalar,Oset-scalar1,Oset-scalar2}. In particular, this
is the case when the resonances in the infinite volume are located close to
the thresholds, so that, in a finite volume, one encounters a difficulty in
separating these two effects in the measured spectrum.  It
has been explicitly demonstrated that, using twisted boundary conditions, it
is possible to move threshold away from the resonance pole position. As a
result, the accuracy of the extracted pole position increases
dramatically~\cite{Oset-scalar1,Oset-scalar2}.  

It should be pointed out that, imposing twisted boundary conditions on the
quark fields implies, in general, calculating gauge configurations
anew. For this reason, the simulations with the so-called ``fully twisted''
quarks are prohibitively expensive. A much cheaper solution that goes under
the name of ``partial twisting,'' uses the same gauge configurations but
twisted valence quarks in the propagators. It is clear that fully and
partially twisted theories differ in a finite volume. Hence, it is legitimate
to ask, whether the spectrum of the partially-twisted theory can be still used
for the calculation of the physical observables.

\begin{figure}[t]
\begin{center}
\includegraphics[width=5.cm]{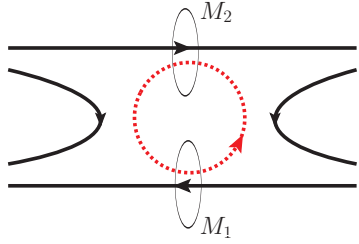}
\end{center}
\caption{An example of an annihilation diagram in meson-meson scattering.
The full and dashed lines denote valence and sea quarks, respectively.
The intermediate state for this diagram consists of two mesons $M_1$ and $M_2$
with one valence and one sea quark.}
\label{fig:annihilation}
\end{figure}

There are the situations, when the use of the partially twisted
boundary conditions can be rigorously justified (see, e.g.,
Refs.~\cite{Sachrajda,Chen}). In particular, these are the situations where
the so-called annihilation diagrams of the type shown in
Fig.~\ref{fig:annihilation} are absent. In this case, it can be proven that
the potential that describes interactions in the system of
 two hadrons is the same in the
infinite and in a finite volume, up to the exponentially suppressed terms.
Consequently, the spectrum in the partially twisted case can be analyzed by
using the L\"uscher equation~\cite{Luescher-torus} which is derived in the fully
twisted theory -- the differences will be exponentially suppressed in a large
volume. 

One may easily see what goes different when the annihilation diagrams like the
one shown in Fig.~\ref{fig:annihilation} are present. In the EFT language, the quark diagram shown in this figure corresponds to the
intermediate state of two fictitious mesons consisting from one valence quark
and one sea antiquark (or {\em vice versa}). The threshold of this diagram
coincides with the elastic threshold -- consequently, the finite-volume
effects in such diagrams are only power-law suppressed and can not be
neglected. On the other hand, including such intermediate states in the
L\"uscher equation explicitly will necessarily lead to a different equation,
since the boundary conditions imposed on the fictitious mesons differ from the
ones imposed on the usual ones (because the boundary conditions on the valence
and sea quarks differ). Consequently, one arrives at the conclusion that, in
the presence of the annihilation diagrams, the equivalence of partial and full
twisting can not be proven, so one either uses full twisting or gives it up.

We consider such a conclusion premature, for the following reason. As it can
be seen from the discussion above, the L\"uscher equation will indeed have to
get modified in the presence of annihilation diagrams. However, such a
modification proceeds in a well-defined manner: only two-particle intermediate
states feel twisting, whereas the interaction potential between various
hadron pairs stays the same in the finite volume. So, the derivation of the
modified L\"uscher equation is a straightforward task. A non-trivial part of
the problem consists in answering the question, whether the modified L\"uscher
equation enables one to extract information about the physical sector of the
theory (i.e., the sector with only valence quarks). If this is the case,
the use of partial twisting can be still justified, even in the presence of
annihilation diagrams.      

In this work, we concentrate on a particular example, namely, the scalar
resonance $a_0(980)$ with a full isospin $I=1$ and provide a complete solution
of the problem in question. From this example it becomes crystal 
clear, how the method would work in a general case. 

\section{Symmetries of the hadronic potential}

A straightforward way to derive L\"uscher equation is to use EFT methods (see,
e.g., Refs.~\cite{Beane,Lage-piN}). It should be stressed that, 
in our case, we speak of {\em two}
effective theories, namely, of the partially twisted Chiral Perturbation Theory (ChPT) 
which is eventually matched to the non-relativistic EFT. The finite-volume
spectrum of the latter theory is described by the L\"uscher equation which we
are aimed at to derive.

In order to arrive at the partially twisted ChPT, a standard procedure can be
used (here we mainly follow Ref.~\cite{Sharpe}). The fermion sector of QCD is
enlarged by introducing the so-called valence, sea and ghost (commuting)
quarks:
\eq
\bar\psi (\not\! D+m)\psi\rightarrow
\bar\psi_{\sf v} (\not\! D+m_{\sf val})\psi_{\sf v}
+\bar\psi_{\sf s} (\not\! D+m_{\sf sea})\psi_{\sf s}
+\bar\psi_{\sf g} (\not\! D+m_{\sf gh})\psi_{\sf g}\, ,
\en
where $m_{\sf val}$, $m_{\sf sea}$, $m_{\sf gh}$ are valence, sea, ghost quark
mass matrices. We take them all equal (unlike the partially quenched
case). However, the $SU(3)$ symmetry is not assumed, $m_u=m_d\neq m_s$ in all
sectors. Partially twisted boundary conditions correspond to imposing twisted
boundary conditions on the valence and ghost quarks and periodic boundary
conditions on the sea quarks.

In the chiral limit, the infinite-volume theory is invariant under the
 graded symmetry group $SU(2N|N)_L\times SU(2N|N)_R\times U(1)_V$, where $N=3$ is the number
of light flavors.
The low-energy effective Lagrangian, corresponding to the case of 
partially twisted boundary conditions, contains the matrix 
$U=\exp\{i\sqrt{2}\Phi/F\}$ of the pseudo-Goldstone fields $\Phi$, 
which transforms under this group as
\eq
U\to LUR^\dagger\, ,\quad\quad L,R\in SU(2N|N)\, .
\en
The Hermitian matrix $\Phi$ has the following representation
\eq
\Phi=
\begin{pmatrix}
{M_{\sf vv} & M_{\sf sv}^\dagger & M_{\sf gv}^\dagger \cr 
M_{\sf sv} & M_{\sf ss} & M_{\sf gs}^\dagger \cr
M_{\sf gv} & M_{\sf gs} & M_{\sf gg}}
\end{pmatrix}\, .
\en 
Here, each of the entries is itself a $N\times N$ matrix in flavor 
space, containing meson fields built up from certain quark species (e.g., from valence quark and valence antiquark, from sea quark and ghost antiquark, and so on).
The fields $M_{\sf gv}$ and $M_{\sf gs}$ are anti-commuting pseudoscalar fields
(ghost mesons). Further, the matrix $\Phi$ obeys the condition
$\mbox{str}\,\Phi=\mbox{tr}\,(M_{\sf vv}+M_{\sf ss}-M_{\sf gg})=0\, ,$
where ``str'' stands for the supertrace.

The effective chiral Lagrangian takes the form
\eq\label{eq:L}
{\cal L}=\frac{F_0^2}{4}\mbox{str}\,(\partial_\mu U\partial^\mu U^\dagger)
-\frac{F_0^2}{4}\,\mbox{str}(\chi U+U\chi^\dagger)+\mbox{higher-order terms,}
\en
where $\chi=2mB_0$ is proportional to the quark mass matrix.

In the infinite volume, the above theory is completely equivalent to ordinary
Chiral Perturbation Theory (ChPT), since the masses of the quarks of all species
are set equal. In a finite volume, the difference arises due to the different
boundary conditions, set on the different meson fields. These boundary 
conditions are uniquely determined by the boundary conditions imposed on the
constituents.

Consider now the S-wave scattering of two pseudoscalar mesons in the channel
with the total isospin $I=1$. This is necessarily a coupled-channel problem,
with the two-particle channels containing mesons from all sectors. These
channels for $I=I_3=1$ are listed in Table~1.

\setcounter{table}{0}
\begin{table}[t]\label{tab:channels}
\begin{eqnarray*}
\begin{array}{|l|l|l|}
\hline\hline
\mbox{Index} & \mbox{Channel} & \mbox{Quark content} \\
\hline
1 & |\pi^+_{\sf vv}\eta_{\sf vv}\rangle & 
-\frac{1}{\sqrt{6}}\,|(u_{\sf v}\bar d_{\sf v})
(u_{\sf v}\bar u_{\sf v}+d_{\sf v}\bar d_{\sf v}-2s_{\sf v}\bar s_{\sf v})\rangle\\
2 & |\pi^+_{\sf vv}\eta'_{\sf vv}\rangle & 
-\frac{1}{\sqrt{3}}\,|(u_{\sf v}\bar d_{\sf v})
(u_{\sf v}\bar u_{\sf v}+d_{\sf v}\bar d_{\sf v}+s_{\sf v}\bar s_{\sf v})\rangle\\
3 & |\pi^+_{\sf vv}\eta_{\sf ss}\rangle & 
-\frac{1}{\sqrt{6}}\,|(u_{\sf v}\bar d_{\sf v})
(u_{\sf s}\bar u_{\sf s}+d_{\sf s}\bar d_{\sf s}-2s_{\sf s}\bar s_{\sf s})\rangle\\
4 & |\pi^+_{\sf vv}\eta'_{\sf ss}\rangle & 
-\frac{1}{\sqrt{3}}\,|(u_{\sf v}\bar d_{\sf v})
(u_{\sf s}\bar u_{\sf s}+d_{\sf s}\bar d_{\sf s}+s_{\sf s}\bar s_{\sf s})\rangle\\
5 & |\pi^+_{\sf vv}\eta_{\sf gg}\rangle & 
-\frac{1}{\sqrt{6}}\,|(u_{\sf v}\bar d_{\sf v})
(u_{\sf g}\bar u_{\sf g}+d_{\sf g}\bar d_{\sf g}-2s_{\sf g}\bar s_{\sf g})\rangle\\
6 & |\pi^+_{\sf vv}\eta'_{\sf gg}\rangle & 
-\frac{1}{\sqrt{3}}\,|(u_{\sf v}\bar d_{\sf v})
(u_{\sf g}\bar u_{\sf g}+d_{\sf g}\bar d_{\sf g}+s_{\sf g}\bar s_{\sf g})\rangle\\
7 & |K^+_{\sf vv}\bar K^0_{\sf vv}\rangle & 
|(u_{\sf v}\bar s_{\sf v})(s_{\sf v}\bar d_{\sf v})\rangle\\
8 & |K^+_{\sf vs}\bar K^0_{\sf vs}\rangle & 
|(u_{\sf v}\bar s_{\sf s})(s_{\sf s}\bar d_{\sf v})\rangle\\
9 & |K^+_{\sf vg}\bar K^0_{\sf vg}\rangle & 
|(u_{\sf v}\bar s_{\sf g})(s_{\sf g}\bar d_{\sf v})\rangle\\
10 & |\pi^+_{\sf vs}\pi^0_{\sf vs}\rangle & 
\frac {1}{2}(-(u_{\sf v}\bar d_{\sf s})(u_{\sf s}\bar u_{\sf v}-d_{\sf s}\bar d_{\sf v})
+(u_{\sf v}\bar u_{\sf s}-d_{\sf v}\bar d_{\sf s})(u_{\sf s}\bar d_{\sf v})\rangle\\
11 & |\pi^+_{\sf vg}\pi^0_{\sf vg}\rangle & 
\frac {1}{2}(-(u_{\sf v}\bar d_{\sf g})(u_{\sf g}\bar u_{\sf v}-d_{\sf g}\bar d_{\sf v})
+(u_{\sf v}\bar u_{\sf g}-d_{\sf v}\bar d_{\sf g})(u_{\sf g}\bar d_{\sf v})\rangle\\
\hline\hline
\end{array}
\end{eqnarray*}
\caption{Scattering channels for the case of $I=I_3=1$.}
\end{table}

At the next step, the partially quenched ChPT is matched to the
non-relativistic EFT {\em with the same hadron spectrum}. The two-particle
scattering amplitude in the non-relativistic EFT obeys the coupled-channel 
Lippmann-Schwinger (LS) equation
\eq
T_{ij}=V_{ij}+\sum_{m,n=1}^{11}V_{im}G_{mn}T_{nj}\, ,\quad\quad
i,j=1,\cdots,11\, .
\en
By using dimensional regularization, the above equation becomes an {\em
  algebraic} equation, where both the $T_{ij}$ and the potential 
$V_{ij}$ are evaluated {\em
  on shell} (the potential $V_{ij}$ 
coincides with the multichannel $K$-matrix in this
formalism). The quantity $G_{ij}$ stands for the two-particle loops in the
intermediate state. This quantity is not diagonal due to the mixing of the
neutral states, so, in order to calculate this quantity, one has first to
diagonalize it the the basis of physical neutral mesons and then use the
prescriptions of the non-relativistic EFT for calculating a loop. The details
can be found in Ref.~\cite{ournew}.

The entries of $T_{ij}$ and $V_{ij}$, corresponding to the scattering fully in
the valence sector, are termed as {\em physical}. The quark diagrams,
describing the amplitudes in this sector, are the same as in ordinary
QCD. However, $T_{ij}$ and $V_{ij}$ contain unphysical entries as well,
describing the transitions between valence/sea/ghost sectors. The quark
diagrams describing these transitions are, in general, different (e.g.,
containing only disconnected contributions). So, the question arises, whether
one is able to relate the finite-volume spectrum of the theory to the physical
matrix elements only.

The key property which allows one to do so is the symmetry of $T_{ij}$ and 
$V_{ij}$, which
stems from the $SU(2N|N)_L\times SU(2N|N)_R\times U(1)_V$ graded 
symmetry of the original theory. In particular, it can be shown that the
matrix elements of these matrices obey certain linear relations which reduce
the number of the independent entries. As a nice check, it can be verified
that, due to these constraints, the relation between the $T$- and $K$-matrix
elements in the infinite volume turns out to be the same as in ordinary ChPT,
without sea and ghost quarks. In fact, in the infinite volume these two
theories should be exactly equivalent.

\section{Derivation of the partially twisted L\"uscher equation}

In a finite volume, only the matrix $G$ containing two-meson loops
changes $G\to G_L$, where the matrix elements of $G_L$ are linear combinations
of the L\"uscher zeta-functions. The potential remains the same. The spectrum is determined
from the secular equation
\eq
\mbox{det}\,(1-VG_L)=0\, .
\en
Various scenarios of the partial twisting lead to the different modifications
of $G$ and hence to the different versions of the L\"uscher equation.
Below, we shall consider two scenarios. The general pattern will be clear from
these examples.

\bigskip
{\bf Scenario 1:}
\medskip

We impose periodic boundary conditions on the $u$-,$d$-quarks and twisted
boundary conditions on the $s$-quark:
\eq
u({\bf x}+{\bf n}L)=u({\bf x})\, ,\quad
d({\bf x}+{\bf n}L)=d({\bf x})\, ,\quad
s({\bf x}+{\bf n}L)=e^{i\theta n}s({\bf x})\, .
\en
These boundary conditions translate into the boundary conditions for the meson
states. Only the boundary conditions for the kaons change:
\eq
K^\pm({\bf x}+{\bf n}L)=e^{\mp i\theta n}K^\pm({\bf x})\, ,\quad\!\!
K^0({\bf x}+{\bf n}L)=e^{-i\theta n}K^0({\bf x})\, ,\quad\!\!
\bar K^0({\bf x}+{\bf n}L)=e^{i\theta n}\bar K^0({\bf x})\, .
\en
This means that $K$ and $\bar K$ mesons {\em containing valence and ghost $s$-quarks} get additional 3-momenta $\mp \theta/L$. The system stays in the CM frame.
The secular equation takes the form
\eq
(1-V_{77} K_L-V_{11} E_L+(V_{77}V_{11}-V_{17}^2)K_L E_L)F_L(\theta)=0\, ,
\en
where $K_L$ and $E_L$ denote the $K\bar K$ and $\pi\eta$ loops in a finite
volume {\em in the absence of twisting} 
\eq
K_L,E_L=\frac{1}{4\pi^{3/2} P_0L}\,Z_{00}(1;q^2)\, ,\quad\quad
q=\frac{pL}{2\pi}\, ,
\en
$p$ is the magnitude of the relative three-momentum of a pair in the CM
frame ($K\bar K$ or $\pi\eta$ pair, respectively), and $P_0$ is the total
energy of a pair. The quantity $F_L(\theta)$ denotes
 a factor that depends on the
unphysical entries of the matrix $V$. 

As seen, owing to the symmetries of the matrix $V$, the determinant in the
secular equation was split, and a piece containing only physical amplitudes,
has emerged. The resulting equation is, however, not very useful because it
coincides with the equation with {\em no twisting}.

As seen, the spectrum of the partially twisted equation contains more states
than the fully twisted one (these are the solutions of the equation
$F_L(\theta)=0$). Physically, these solutions are not interesting because the
physical and non-physical matrix elements are intertwined here. 
It could be shown that, choosing particular source/sink operators, which do
not have an overlap with some of the states, one may project out the physical part of
the spectrum.

\bigskip
{\bf Scenario 2:}
\medskip

Here we consider twisting of the $u$-quark, leaving
the $d$- and $s$-quarks to obey periodic boundary conditions. What changes
here is the free Green function in a finite volume.
\eq
K,E\to K_L^\theta,E_L^\theta
=\frac{1}{4\pi^{3/2}\sqrt{s}\gamma L}\,Z_{00}^{\bf d}(1;(q^*)^2)\, ,\quad
q^*=\frac{p^*L}{2\pi}\, ,\quad
 p^*=\frac{\lambda^{1/2}(s,m_1^2,m_2^2)}{2\sqrt{s}}\, .
\en
where ${\bf d}={\bf P}L/2\pi=\theta/2\pi$,
$s=P_0^2-{\bf P}^2$, $\gamma=P_0/\sqrt{s}$, 
the quantity $\lambda(x,y,z)$ is Mandelstam triangle function
and $Z_{00}^{\bf d}(1;(q^*)^2)$ denotes the L\"uscher zeta-function in the moving frame~\cite{Rummukainen}, see also Refs.~\cite{Schierholz,Hoja}:
\eq
Z_{00}^{\bf d}(1;(q^*)^2)&=&\frac{1}{\sqrt{4\pi}}\,\sum_{{\bf r}\in P_d}
\frac{1}{{\bf r}^2-(q^*)^2}\, ,
\nonumber\\[2mm]
P_d&=&\{{\bf r}=\mathbb{R}^3~|~r_\parallel=\gamma^{-1}(n_\parallel-\mu_1|{\bf d}|),
~{\bf r}_\perp={\bf n}_\perp,~{\bf n}\in\mathbb{Z}^3\}\, ,
\en
where $\mu_1=\bigl(1-(m_1^2-m_2^2)/s\bigr)/2$.

The secular equation in a finite volume takes the form
\eq
(1-V_{77} K_L^\theta-V_{11} E_L^\theta+(V_{77}V_{11}-V_{17}^2)K_L^\theta E_L^\theta)F'_L(\theta)=0\, ,
\en
where $F'_L(\theta)$ is another factor, depending on the unphysical entries.
It is seen that the spectra in case of the partial and full twisting coincide.

\section{Summary}

Using the non-relativistic EFT technique in a finite volume, we have derived 
the L\"uscher equation for the partially twisted boundary conditions for
coupled-channel $\pi\eta-K\bar K$ scattering.
Our final result is remarkably simple. If in the channel with $I=I_3=1$
the light quarks are subject to twisting, the partially twisted L\"uscher
equation is equivalent to the fully twisted one, despite the presence
of annihilation diagrams. If, on the contrary, partial twisting of the
 strange quark is performed, the physically interesting part of the 
spectrum is not affected. Other scenarios are also possible and can be 
investigated by using the same methods. 

We think that this result would be
interesting for the lattice practitioners studying the properties of
scalar mesons. We have shown that, instead of carrying out simulations at
different volumes, as required in the L\"uscher approach, one may perform
relatively cheaper partially twisted simulations.

\section*{Acknowledgments} 
The authors thank S. Beane, J. Bijnens, J. Gasser, 
T. L\"ahde, Ch. Liu, M. Savage, S. Sharpe
and C. Urbach for interesting discussions. 
One of us (AR) thanks the 
Institute for Nuclear Theory at the University of Washington 
for its hospitality and the Department of Energy for partial support 
during the completion of this work.
This work is partly supported by the EU
Integrated Infrastructure Initiative HadronPhysics3 Project  under Grant
Agreement no. 283286. We also acknowledge the support by the DFG (CRC 16,
``Subnuclear Structure of Matter''), by the DFG and NSFC 
(CRC 110, ``Symmetries and the Emergence
of Structure in QCD''), by
the Shota Rustaveli National Science Foundation
(Project DI/13/02) 
and by the Bonn-Cologne Graduate School of Physics and Astronomy.
This research is supported in part by Volkswagenstiftung
under contract no. 86260.

\end{document}